\documentclass[12pt,preprint]{aastex}

\usepackage{natbib}
\slugcomment{Submitted to Astrophys J.}

\shorttitle{Dynamics and neutrino signal of black hole formation}
\shortauthors{Sumiyoshi et al.}

\begin{document}


\title{Dynamics and neutrino signal of black hole formation \\
in non-rotating failed supernovae. II. progenitor dependence}


\author{K. Sumiyoshi}
\affil{Numazu College of Technology, 
       Ooka 3600, Numazu, Shizuoka 410-8501, Japan}
\email{sumi@numazu-ct.ac.jp}

\author{S. Yamada}
\affil{Science and Engineering
       \& 
       Advanced Research Institute for Science and Engineering, \\
       Waseda University, 
       Okubo, 3-4-1, Shinjuku, Tokyo 169-8555, Japan}

\and

\author{H. Suzuki}
\affil{Faculty of Science and Technology, Tokyo University of Science,\\
       Yamazaki 2641, Noda, Chiba 278-8510, Japan}




\begin{abstract}
We study the progenitor dependence of the black hole formation 
and its associated neutrino signals from the gravitational collapse 
of non-rotating massive stars, 
following the preceding study on the single progenitor model in \citet{sum07}.  
We aim to clarify whether the dynamical evolution toward 
the black hole formation occurs in the same manner for different progenitors 
and to examine whether the characteristic of neutrino bursts is general 
having the short duration and the rapidly increasing average energies.
We perform the numerical simulations by general relativistic 
$\nu$-radiation hydrodynamics to follow the dynamical evolution 
from the collapse of pre-supernova models of 40M$_{\odot}$ and 
50M$_{\odot}$ toward the black hole formation via contracting 
proto-neutron stars.  
For the three progenitor models studied in this paper, 
we found that the black hole formation occurs in $\sim$0.4--1.5 s 
after core bounce through the increase of proto-neutron star mass 
together with the short and energetic neutrino burst.  
We found that density profile of progenitor is important 
to determine the accretion rate onto the proto-neutron star 
and, therefore, the duration of neutrino burst.  
We compare the neutrino bursts of black hole forming events 
from different progenitors and discuss whether we can probe 
clearly the progenitor and/or the dense matter.  
\end{abstract}



\section{Introduction}\label{sec:intro}

The final fate of massive stars is one of key issues in stellar 
physics as well as astrophysics.  
The massive stars beyond $\sim$8M$_{\odot}$ are the origin of compact 
objects, i.e. neutron stars and black holes, as the outcome of 
gravitational collapse at the end of their lives.
They contribute also to the production of heavy elements through 
the explosive nucleosynthesis and to the high energy phenomena 
with the bursts of electro-magnetic radiations and neutrinos, 
which in turn play important roles in the evolution of galaxies.  
The massive stars in the wide range between $\sim$8M$_{\odot}$ and 
$\sim$100M$_{\odot}$ lead to various explosive phenomena such as 
core-collapse supernovae, hypernovae (or gamma ray bursts) and 
failed supernovae \citep{heg03}.  
Such energetic phenomena are dependent on the properties of 
progenitors such as masses, rotations and profiles, 
for example, as 
in the mass limit of ordinary supernovae and the branches of 
hypernovae and failed supernovae as a function of mass \citep{mae03}.
We focus here on 
the fate of non-rotating massive stars, 
as an extreme of failed supernovae, in the mass range 
of $\sim$40--50M$_{\odot}$.  

Such a death of massive stars attracts attention recently 
since it is one of possible channels of the black hole formation and 
is characterized by the unique neutrino signals \citep{lie04,sum06,sum07}:  
the short duration and increase of average energies and luminosities 
in time. 
This is in contrast to the ordinary supernova neutrinos, 
whose emissions from a nascent proto-neutron star 
last for $\sim$20 s with a gradual decrease of average energies 
and luminosities \citep{bur88,suz94}.  
The black hole formation will be inevitable for massive stars 
beyond $\sim$40M$_{\odot}$, which was estimated with large
uncertainties to be the mass threshold for black hole formation 
by \citet{fry99}. 
The dynamical collapse to a black hole is triggered 
by the intense accretion of matter onto the proto-neutron star.  
Note that this is different from the so-called delayed black hole formation 
of the proto-neutron star with a fixed mass \citep{kei95,pon99,bau96b}
owing to the termination of accretion after a successful launch of shock wave.
It should be also emphasized that the neutrino 
signal may be used as a probe into the equation of state of nuclear 
matter \citep{sum06} and exotic matter \citep{nak08} in a different way 
from the delayed collapse of proto-neutron stars \citep{pon01a,pon01b}.  

This article is the second paper, 
following the first paper \citep{sum07}, 
of serial studies on the black hole formation associated 
with the short burst of energetic neutrinos.  
In the first paper, 
the evolution toward the black hole formation 
from the gravitational collapse of the 40M$_{\odot}$ star
have been examined by taking account of 
the dependence of equation of state.  
By solving general relativistic $\nu$-radiation hydrodynamics, 
they have followed the hydrodynamics up to the moment of black hole formation 
and the neutrino distributions during the whole evolution 
to obtain the detailed features of neutrino emission.  
They have shown that the black hole formation occurs in $\sim$1 s 
after the core bounce via proto-neutron star evolution, 
but it depends crucially on the equation of state of dense matter.  
The equation of state (EOS) controls the timing of the re-collapse 
from the proto-neutron star to the black hole since the EOS determines 
the maximum mass.  
The mass of proto-neutron star increases due to the accretion of 
material in the situation of failed shock wave and the dynamical 
collapse is triggered when the mass reaches the critical mass.
We remark that the progenitor model of 40M$_{\odot}$ by \citet{woo95} 
has been uniquely adopted in their studies and 
other studies on such massive stars \citep{lie04}.  

In the first paper \citep{sum07}, 
associated neutrino burst has been shown to be 
unique to identify the black hole formation and to probe 
the properties of dense matter.  
The neutrino burst is terminated by the black hole formation 
and the average energies of neutrinos increase toward the end 
because of high temperature realized in the collapsing proto-neutron stars.  
These features are again sensitive to the properties of equation 
of state, which determines the duration of neutrino burst and 
the properties of compressed matter at neutrino emitting region.  
The neutrino burst toward the black hole formation can, in principle, 
be detectable at the terrestrial neutrino detector facilities 
\citep{ike07} 
and can be used as a signature of black hole formation and 
a probe of dense matter \citep{sum06}.  
Moreover, if a collapse of massive star occurs in a nearby
galaxy, the event may be observed, for example, in the proposed survey of 
disappearance of massive stars \citep{koc08} and  
it will be possible to obtain the information on the progenitor.  
However, one has to provide the detailed features of neutrino burst 
by taking care of uncertainties of progenitor models in addition 
to dense matter if one wants to use them as templates for the 
neutrino search.  

In this study, we 
make a first attempt to see 
whether the features of black hole 
formation reported in the first paper \citep{sum07} 
are taken over to 
different progenitors.  
We adopt 
three 
stellar models as the initial condition of 
the numerical simulations to make the comparison with the evolutions 
by the numerical modeling we have adopted in the previous study \citep{sum07}.
In addition to the stellar model of 40M$_{\odot}$ star by \citet{woo95}, 
we adopt the two stellar models by \citet{has95} and by \citet{tom07}.
As a case of different mass,
we adopt a 50M$_{\odot}$ star by \citet{tom07}, who studied 
the evolution of metal-poor massive stars.  
We have chosen the model with zero metallicity 
among their series of different metallicities.
This model is the limiting case of metal-poor massive stars 
without mass loss.  
As a model for the same mass (40M$_{\odot}$) but by a different method 
of stellar evolutionary calculation, 
we adopt a 40M$_{\odot}$ star by \citet{has95}.  
The comparison between the different models enables us to explore 
the dependence of numerical results on the uncertainty 
in the stellar evolution.  

The investigation on the progenitor dependence is important 
for the utilization of neutrino bursts as a probe of equation of state.
If the neutrino burst is quite different depending on the progenitors, 
it becomes difficult to distinguish the differences found by the equation 
of state.  
Using the two sets of EOS (LS-EOS and Shen-EOS), it has been shown that 
the time duration of neutrino bursts differ by a factor of $\sim$2 depending 
on the softness of EOS.  
If the progenitor models give large differences, they may smear out 
the EOS difference, which we want to use as a probe.

Our aim of the paper is, therefore, to 
make a first exploration to 
the dependence 
on the progenitors by adopting 
a small number of 
stellar models 
and to show that the dynamics toward the black hole formation 
is common in the massive stars having a large central iron core.  
We aim also to demonstrate that the neutrino signal has similar 
features once we set the equation of state.  
We analyze how the evolution is related with the density 
profile of progenitor, which determines the accretion rate.  
We discuss the characteristics of neutrino bursts obtained by different 
progenitors with the sets of equation of state 
for possible detection in future.  

We plan this paper as follows.  
We briefly explain the method of numerical simulations 
in \S \ref{sec:simulations}.  
We compare the progenitors adopted for initial models 
and discuss the differences in \S \ref{sec:initial}.  
We report the numerical results on the dynamics 
in \S\S \ref{sec:50M} and \ref{sec:40M}
for two models through the comparison with the previous result.
We describe the accretion rates and the profiles during the 
evolution in \S\S \ref{sec:accretion} and \ref{sec:profile} 
to explain the differences among the models.  
We discuss the characteristics of neutrino signals in the models 
to distinguish the progenitor and the equation of state 
in \S \ref{sec:nu-signal}.  
Implications of this work and summary will be made 
in \S\S \ref{sec:discussions} and \ref{sec:summary}.  

\section{Numerical Simulations}\label{sec:simulations}

Numerical simulations are performed by the numerical code 
of general relativistic $\nu$-radiation hydrodynamics 
under the spherical symmetry \citep{yam97,yam99,sum05}.
This numerical code has been applied to the gravitational 
collapse of massive stars to study the supernova explosion \citep{sum05} 
and the black hole formation \citep{sum06,sum07,nak06,nak07}.  
The detailed description on the numerical simulations 
of the gravitational collapse toward the black hole formation 
can be found in \citet{sum07}, which reported also the results 
for the progenitor of a 40M$_{\odot}$ star by \citet{woo95}.
We adopt the same number of grids for lagrangian mass coordinate, 
variables in neutrino distributions for $\nu_e$, $\bar{\nu}_e$, 
$\nu_{\mu/\tau}$ and $\bar{\nu}_{\mu/\tau}$.  
The mesh for mass coordinate is optimized for each progenitor model 
to follow the accretion phase with an enough resolution 
using the rezoning method.  

The fully implicit method of $\nu$-radiation hydrodynamics 
enables us to follow the long evolution over $\sim$1 s 
toward the black hole formation.  
The exact treatment of $\nu$-transfer together with hydrodynamics 
in general relativity permits the detailed evaluations of the 
neutrino fluxes from the contracting proto-neutron stars 
until the black hole formation.  
We define the black hole formation by finding the apparent horizon 
as explained in \citet{nak06} and \citet{sum07}.  

We adopt the microphysics in the same manner as in \citet{sum07} 
to facilitate the comparison of new results with the previous 
result for 40M$_{\odot}$.  
We use the two sets of equation of state (EOS) for supernova simulations
to assess the influence of EOS on different progenitors.
The EOS by \citet{she98a,she98b} (Shen-EOS), which is obtained 
by the relativistic mean field framework with the data of unstable nuclei, provides a rather stiff EOS 
having the maximum mass of neutron stars to be 2.2M$_\odot$.
The EOS by \citet{lat91} (LS-EOS) is obtained by the 
extension of the compressible liquid model, taking a density 
dependence of energy based on the Skyrme-interaction.
We have chosen the case of incompressibility of 180 MeV among the three 
choices in LS-EOS as a representative of soft EOS having the 
maximum mass of neutron stars to be 1.8M$_\odot$.  
The same set of weak interaction rates as in \citet{sum07} 
is adopted for the current simulations by implementing 
the standard formulation of \citet{bru85} 
and the several extensions \citep{sum05}.

\section{Initial Models}\label{sec:initial}

We adopt three different profiles from the pre-supernova models 
of massive stars as initial models.  
We have picked massive stars around 40M$_{\odot}$, 
which are massive enough to have the black hole formation 
without the successful explosion.  
The choice of the progenitor models in the current study 
is meant for a first and small-scale trial exploration and 
does not reflect a possible great variety of massive stars.  
We use the central part of pre-supernova profiles up to 3M$_{\odot}$ 
in baryon mass coordinate and set density, electron fraction and 
temperature distributions as initial configurations 
for numerical simulations.  

As a base line model, we adopt a 40M$_{\odot}$ star 
from the set of massive stars by \citet{woo95}.  
This model has been used for the previous calculations of 
our studies of black hole formation and the details of 
numerical results can be found in \citet{sum07}.  
We name this model W40 and will use as a reference 
to examine the new results.  
The size of iron core in W40 model is 1.98M$_{\odot}$.  
As an example of other types of massive stars, we adopt 
a case of 50M$_{\odot}$ with zero metallicity (Z=0) 
from the series of massive stars 
having low metallicities \citep{ume05,tom07}.  
These zero-metal stars do not 
experience the mass loss during the stellar evolution 
and remain massive as in the initial stage \citep{ume00}.  
This choice is meant to explore the dependence 
of the phenomena on the progenitor mass and the stellar 
evolution.  
The size of iron core is 1.88M$_{\odot}$.\footnote{This value 
is determined by the criteria of $Y_{e} < 0.50$, 
which corresponds roughly to the $^{56}$Ni abundant region.  
A smaller value, 1.46M$_{\odot}$, is obtained by the criteria of 
$Y_{e} < 0.49$, which is the convention in their study 
(N. Tominaga, private communication, 2006). }
We name this model T50.
In order to examine the difference due to the modeling 
of stellar evolutions for the same stellar mass, 
we adopt a 40M$_{\odot}$ star by \citet{has95}.
The size of iron core in this model is 1.88M$_{\odot}$, 
which is largest among the series of his models up to 70M$_{\odot}$.  
We name this model H40.  

\begin{figure}
\epsscale{.70}
\plotone{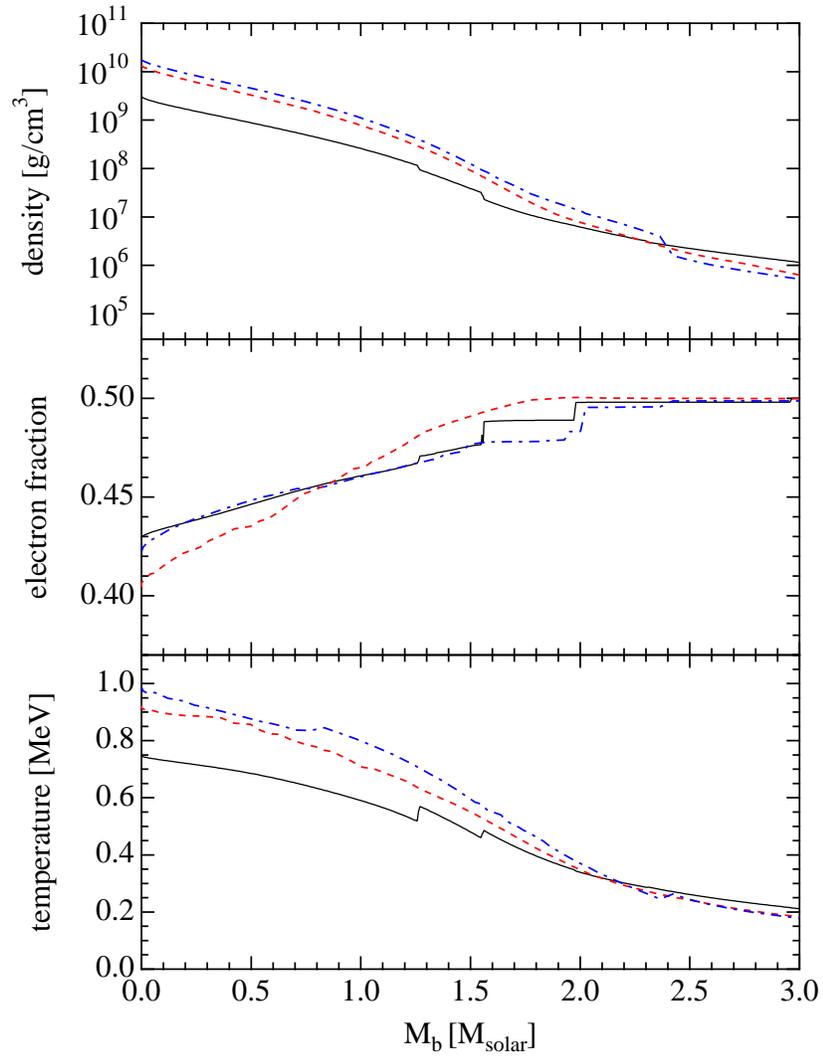}
\caption{Density, electron fraction and temperature distributions 
of progenitor models are shown as a function of baryon mass coordinate 
by solid, dashed and dash-dotted lines for W40, T50 and H40, respectively.}
\label{fig:progenitor}
\end{figure}

We compare the profiles of the three adopted models 
in Fig.~\ref{fig:progenitor}.  
We can see quantitative differences among the three models 
though the profiles look similar at first glance.  
The density of T50 and H40 are higher than 
those of W40 in the central region 
whereas the order is reversed at the outer region.
This difference in gradients of density profile 
from the center to the outer part 
is related to the difference in accretion rates onto proto-neutron star.  
Actually, the densities at M$_b$ = 2--3M$_{\odot}$, 
which cover the range for the maximum mass of proto-neutron stars, 
are crucial to determine the final stage of accreting proto-neutron star 
through the free-fall time scale ($\sim 1 / \sqrt{G \rho}$).  

The profile of electron fraction in T50 is slightly different 
from those in W40 and H40.
There are several kinks at different positions due to the shell 
structure.  
The temperature of T50 and H40 are higher than that of W40 
in the central region. 
This difference, however, smears out during the collapse 
when we compare the profiles at the timing 
with the same central density.  
We will discuss the profiles later in \S \ref{sec:profile} 
to explain the behavior of accretion rates.  

\section{Numerical Results}\label{sec:results}

We present the numerical results for the cases 
of T50 (\S \ref{sec:50M}) and H40 (\S \ref{sec:40M}) 
through the comparison with the base line model of W40.  
We add a character, S or L for Shen-EOS or LS-EOS, respectively, 
in the model names to denote the choice of the EOS.
We summarize the calculated models in Table~\ref{tb:model}.
We discuss the differences among the three models and 
their implications in \S\S \ref{sec:accretion}--\ref{sec:nu-signal}.  

\subsection{50M$_{\odot}$ star by Tominaga et al.}\label{sec:50M}

\begin{figure}
\epsscale{.70}
\plotone{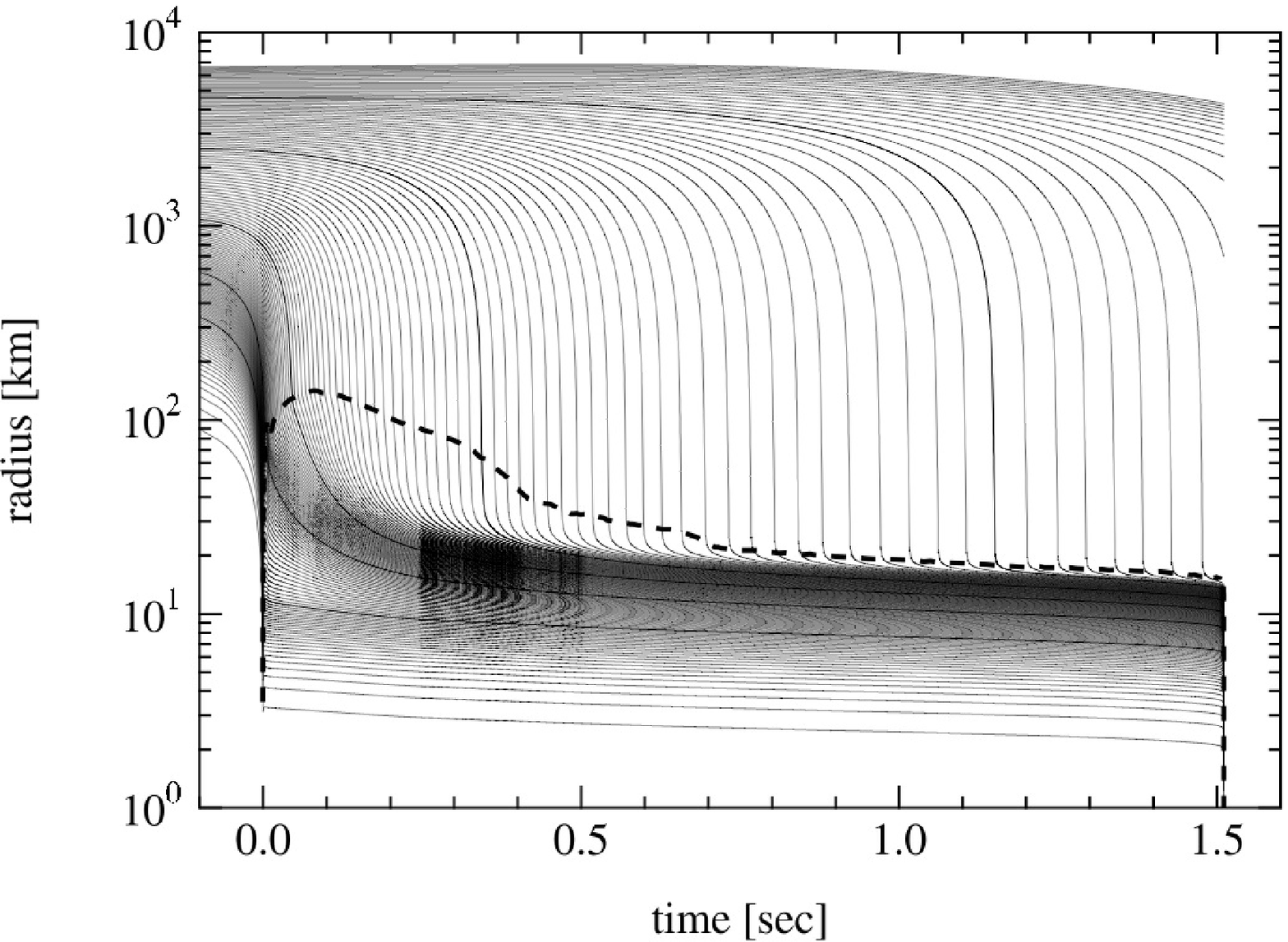}
\caption{Radial trajectories of mass elements 
of the core of 50M$_{\odot}$ star 
as a function of time after bounce in model T50S.
The location of shock wave is displayed by a thick dashed line.}
\label{fig:traj-rhr50t01}
\end{figure}

\begin{figure}
\epsscale{.38}
\plotone{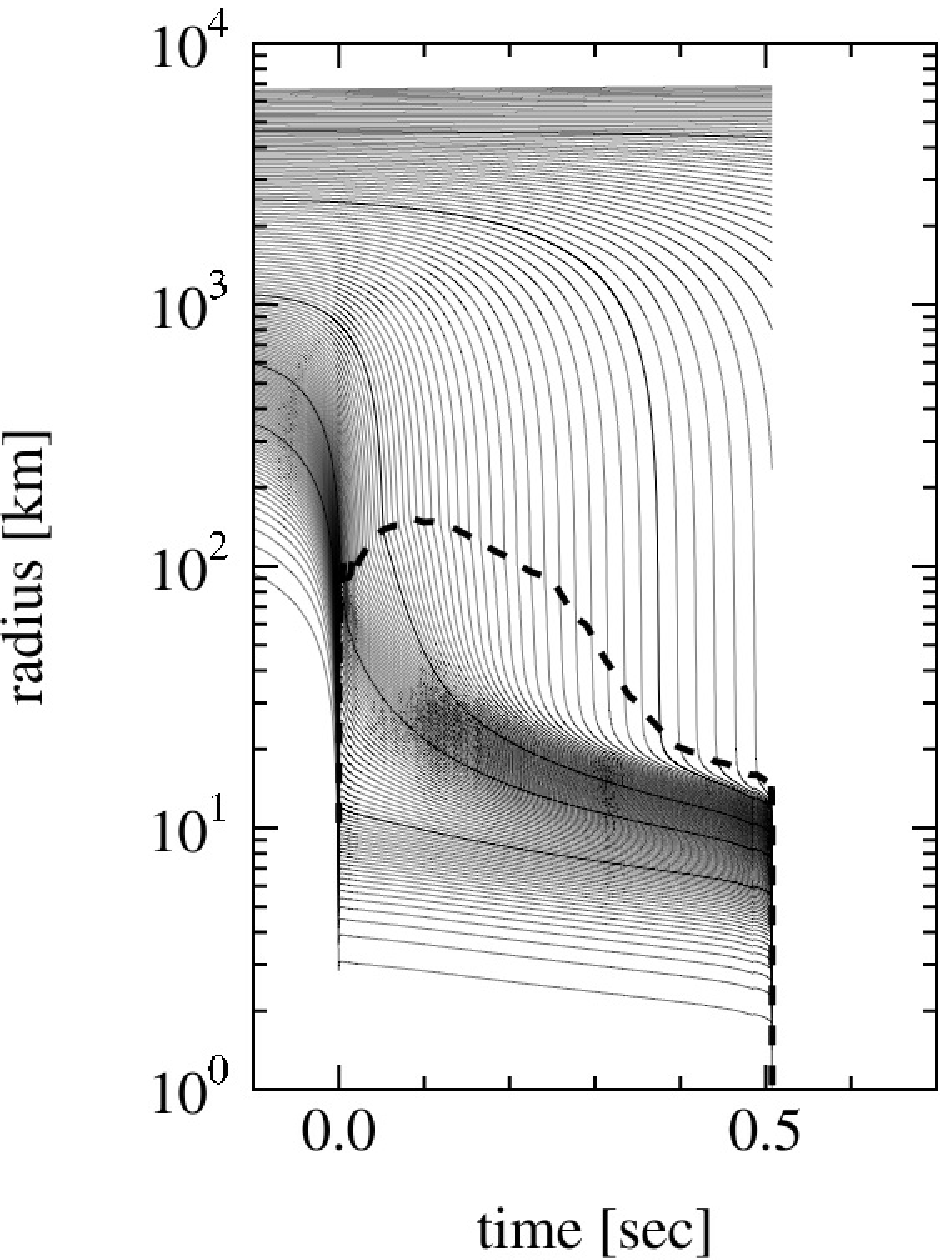}
\caption{Radial trajectories of mass elements 
of the core of 50M$_{\odot}$ star 
as a function of time after bounce in model T50L.
The location of shock wave is displayed by a thick dashed line.}
\label{fig:traj-lhr50t01}
\end{figure}

We show the radial trajectories of mass elements 
as a function of time after bounce for models with T50S and T50L 
in Figs.~\ref{fig:traj-rhr50t01} and \ref{fig:traj-lhr50t01}, respectively.  
The trajectories are plotted for each 0.02M$_{\odot}$ in mass coordinate.  
Thick lines denote the trajectories for 0.5M$_{\odot}$, 1.0M$_{\odot}$, 
1.5M$_{\odot}$, 2.0M$_{\odot}$ and 2.5M$_{\odot}$.  
Thick dashed line represents the position of shock wave.  

The general feature of evolution is found to similar to 
the case of corresponding W40 models.  
After the core bounce, the shock wave is launched up to $\sim$150 km 
in 0.1 s and then recedes gradually toward the surface of central object.  
The nascent proto-neutron star shrinks due to the mass increase 
slowly in T50S (quickly in T50L) by the intense accretion.
The dynamical collapse occurs when the proto-neutron star mass 
exceeds a critical mass, which depends on the stiffness of EOS, 
and the black hole is formed.  

The duration from the core bounce to the black hole formation is 
1.51 s and 0.51 s for T50S and T50L, respectively, 
reflecting the difference between SH-EOS and LS-EOS.  
The duration in T50S is slightly longer than that in W40S, 
while the duration in T50L is slightly shorter than that in W40L.  
This is because the accretion rate is different between T50 and W40 
in a time-dependent manner as we will see in \S \ref{sec:accretion}.  

\begin{figure}
\epsscale{0.8}
\plotone{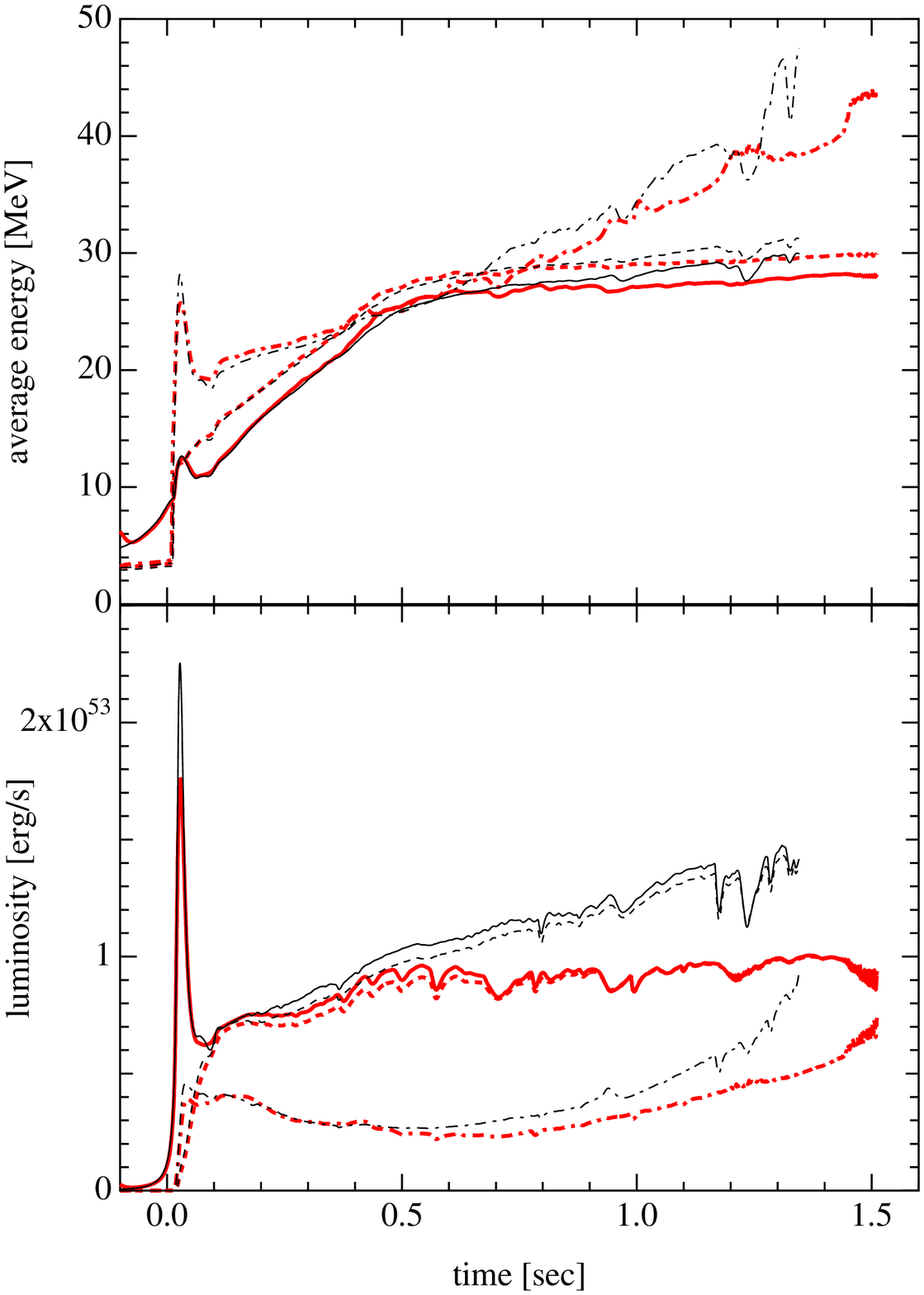}
\caption{Average energies and luminosities 
of $\nu_e$ (solid), $\bar{\nu}_e$ (dashed) and $\nu_{\mu/\tau}$ (dash-dotted) 
for model T50S are shown as a function of time after bounce.  
The results for model W40S are shown by thin lines.  }
\label{fig:EnuLnu-T50S}
\end{figure}

\begin{figure}
\epsscale{0.47}
\plotone{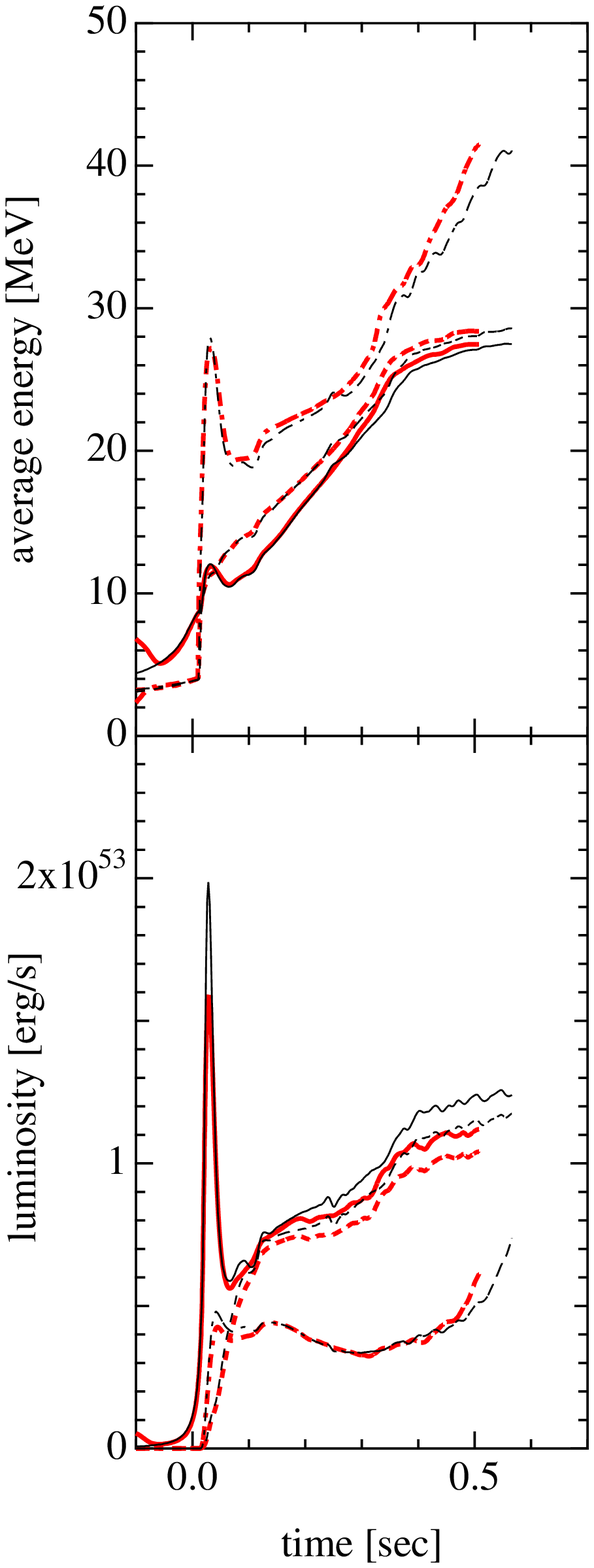}
\caption{Average energies and luminosities 
of $\nu_e$ (solid), $\bar{\nu}_e$ (dashed) and $\nu_{\mu/\tau}$ (dash-dotted) 
for model T50L are shown as a function of time after bounce.  
The results for model W40L are shown by thin lines.  }
\label{fig:EnuLnu-T50L}
\end{figure}

The duration of neutrino bursts is determined by the time 
toward the black hole formation.  
In Figs.~\ref{fig:EnuLnu-T50S} and \ref{fig:EnuLnu-T50L}, 
we display the time evolution of average energies and 
luminosities of neutrinos emitted during the evolution 
described above.  
The corresponding results with the same EOS 
but using the progenitor of W40 are shown by thin lines 
in order to see the progenitor dependence.
In general, the features of neutrino emission for T50 
resemble those for W40.  
The average energies for all species show the increasing 
tendency after bounce.  
The persistent increase for $\nu_{\mu/\tau}$ is remarkable 
among them.  
Looking more closely, small differences appear 
at late phase (after $\sim$0.4 s).  
The luminosities for T50S become clearly lower than 
those for W40S at the late phase.  
This is related with a lower accretion rate in T50 
than that in W40 at $\sim$1 s after bounce (see Fig.~\ref{fig:protoNSmass}).  

We remark that a rapid decrease of neutrino energies 
by the gravitational redshift that is expected at the terminal phase has not
been observed yet in the current simulations.  
One has to follow the evolution further ($\lesssim$20 ms) after the
formation of the apparent horizon.  
For that purpose, however, a
singularity avoiding scheme \citep{bau96a} should be implemented.  
We note also that some wiggles and/or jumps 
seen in Figs.~\ref{fig:EnuLnu-T50S} and \ref{fig:EnuLnu-T50L} 
are numerical artifacts caused by insufficient numerical 
resolutions at the final phase of numerical simulations.  

%

\subsection{40M$_{\odot}$ star by Hashimoto}\label{sec:40M}

\begin{figure}
\epsscale{0.38}
\plotone{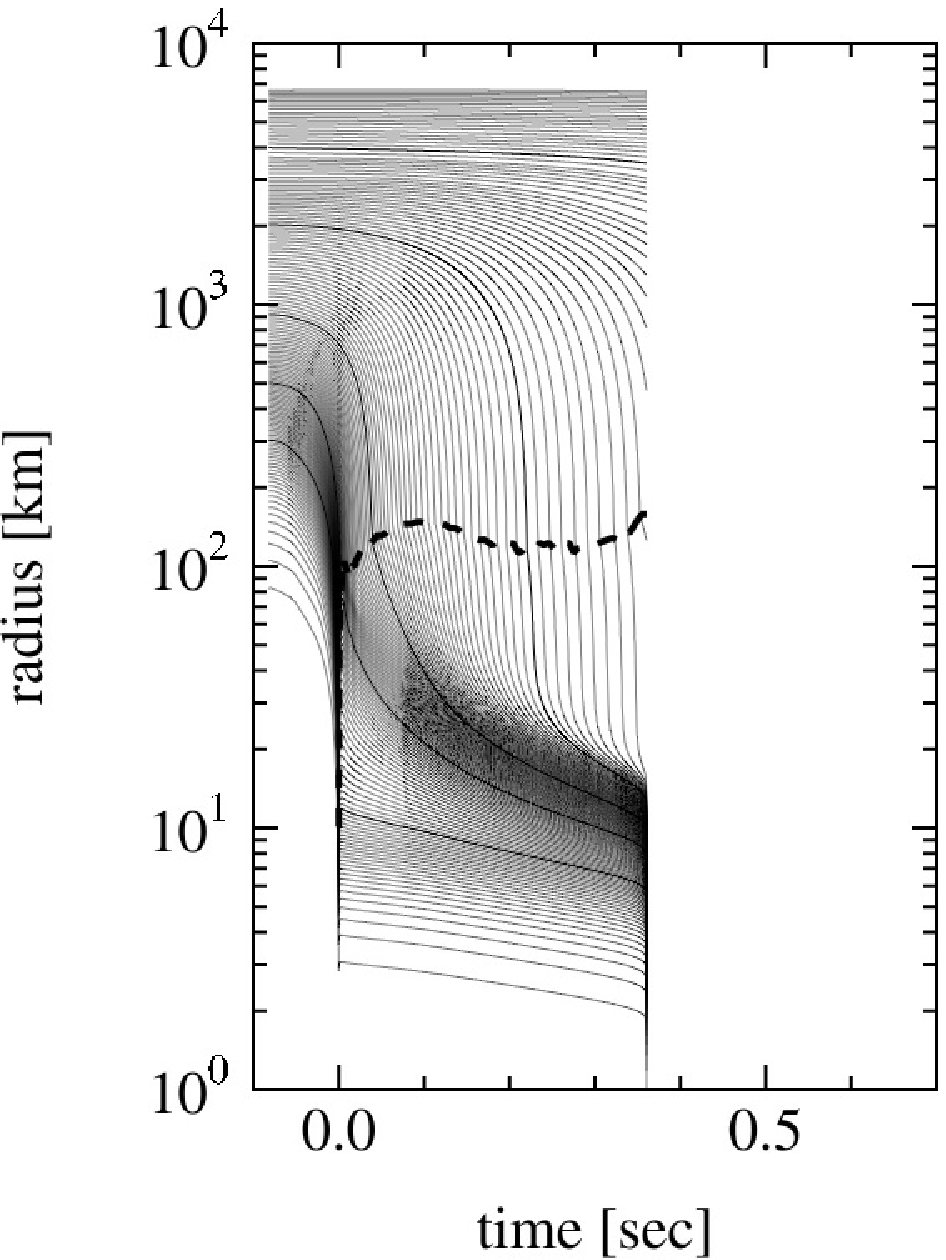}
\caption{Radial trajectories of mass elements 
of the core of 40M$_{\odot}$ star 
as a function of time after bounce in model H40L.
The location of shock wave is displayed by a thick dashed line.}
\label{fig:traj-lhr40h01}
\end{figure}

We display the radial trajectories of mass elements 
as a function of time after bounce for model with H40L 
in Fig.~\ref{fig:traj-lhr40h01}.  
The notation is the same as in Fig.~\ref{fig:traj-rhr50t01}.  
In this case, the evolution is much quicker than 
the cases in W40L and T50L.  
The time till the black hole formation is 0.36 s in H40L.  
Since the accretion rate in H40 is larger than 
those in W40 and T50, 
the mass of proto-neutron star reaches the critical mass 
earlier (see \S \ref{sec:accretion}).  
The contraction of central object is quick accordingly 
and, therefore, the density and temperature inside become 
high early on.  
It is recognizable that 
the position of shock wave behaves in a different manner 
from the other cases.    
The shock wave stays around 100 km after the initial launch 
instead of the recession.  
This is mainly because the neutrino luminosity is high 
due to the high accretion rate.  

\begin{figure}
\epsscale{0.47}
\plotone{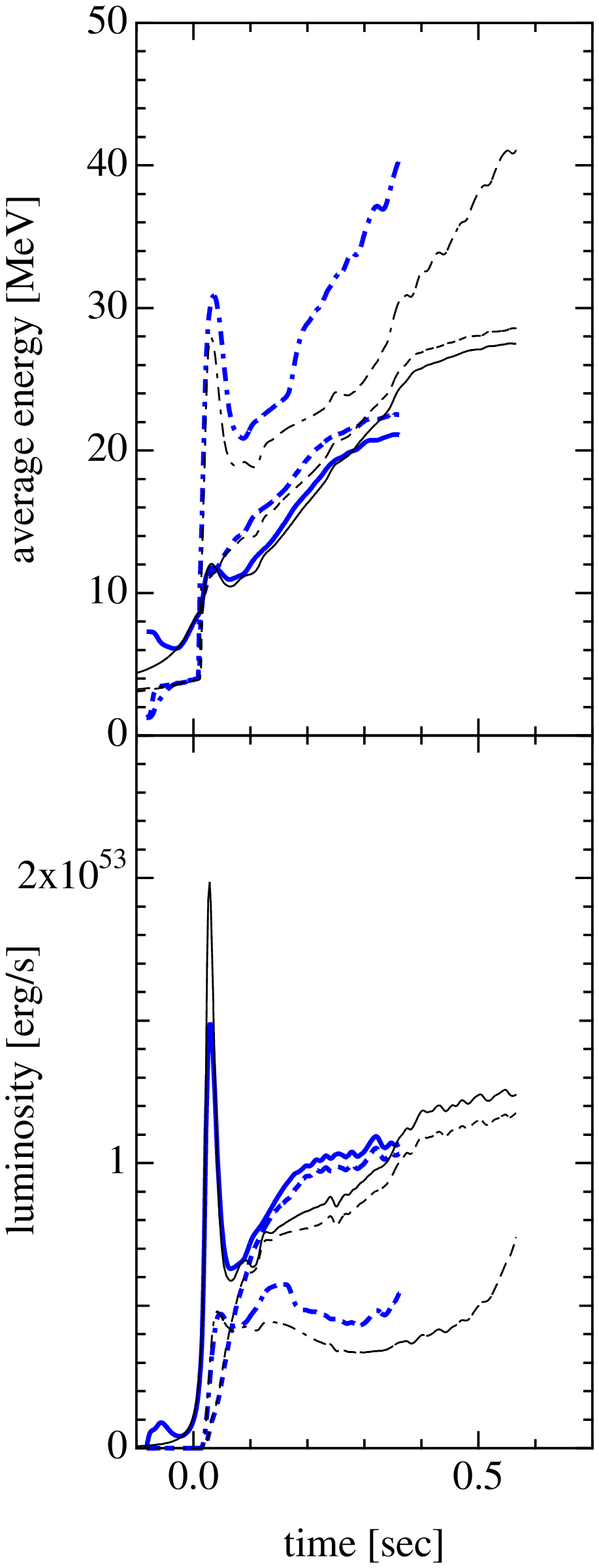}
\caption{Average energies and luminosities 
of $\nu_e$ (solid), $\bar{\nu}_e$ (dashed) and $\nu_{\mu/\tau}$ (dash-dotted) 
for model H40L are shown as a function of time after bounce.  
The results for model W40L are shown by thin lines.  }
\label{fig:EnuLnu-H40L}
\end{figure}

We show the time evolution of average energies and 
luminosities of neutrinos in Fig.~\ref{fig:EnuLnu-H40L}.  
The duration of neutrino burst is clearly shorter than 
that of W40L.  
The average energy of $\nu_{\mu/\tau}$ is larger 
and increases much faster than that of W40L, 
reflecting the different temperatures, 
whereas those of $\nu_e$ and $\bar{\nu}_e$ behave 
similarly.  
The neutrino luminosities of all species in H40L 
are higher than those in W40L.  
The luminosities of $\nu_e$ and $\bar{\nu}_e$ 
are proportional to  the accretion luminosity 
($\sim GM\dot{M}/r$) and are high due to a high 
accretion rate ($\dot{M}$) and 
a more massive object ($M$) at center.  
The high luminosity of $\nu_{\mu/\tau}$ originates 
from a high temperature in the massive object.  
The fact that the luminosity reflects 
the behavior of the accretion luminosity was also seen 
for the neutrino emissions at the shock breakout in the  
core-collapse of 11--20M$_{\odot}$ stars \citep{tho03}.
The larger accretion rates found in the current study of 
40--50M$_{\odot}$ stars lead to more rapid increase of 
luminosity than for less massive stars.  

\subsection{Accretion rate}\label{sec:accretion}

As we have discussed already, the accretion rate is 
a key factor to determine the behavior toward the black hole formation.  
In other words, the increasing rate of proto-neutron star mass 
is crucial to determine 
the growth time to the critical mass.  
We show in Fig.~\ref{fig:protoNSmass} the baryon mass 
of central object (corresponding to the proto-neutron star) 
as a function of time after bounce.  
We define here the central object, which is quasi-hydrostatic, 
as the one inside the position of shock wave.  
Note that the time derivative of the mass curve (i.e. the 
gradient of the curve) is the accretion rate.  
After the core bounce, 
the central object acquires $\sim$1.5M$_{\odot}$ within 50 ms.  
The proto-neutron star with a typical mass is formed 
immediately after the core bounce.
This feature is common for all models of black hole forming 
collapse and is different from the case in ordinary supernovae, 
where the proto-neutron star ($\sim$1.5M$_{\odot}$) formation
takes place in $\sim$0.3 s 
\citep{bur88,suz94}.  
The mass increases gradually toward the critical mass in each 
model and the dynamical collapse to black hole occurs 
at the end point.
The end point depends on the adopted progenitor and 
the equation of state.  

Amazingly, the increase of mass in the cases of W40 and T50 
are similar each other.  
Although the masses for T50 are slightly larger 
up to 0.6 s after bounce, their gradients (i.e. accretion rate) 
are the same, judging from the parallel curves between T50S and W40S 
(T50L and W40L).  
This similarity is the reason why the evolution of T50 and W40 
and resulting neutrino signals are quite similar.  
Adopting LS-EOS, the black hole formation occurs around the same 
time ($\sim$0.5 s) in the cases of T50L and W40L.  
In the case of SH-EOS, however, there is a difference of about 
0.2 s between the cases of T50S and W40S.  
This is due to a smaller accretion rate in T50 at late phase 
($\gtrsim$0.6 s after bounce) than that in W40 as seen in the 
crossing of curves for T50S and W40S.  
We note that 
the small accretion rate at the late phase is related with the 
density profile of outer layer in the adopted progenitor.  

It is remarkable that the behavior of H40L is different.  
The mass of central object for H40L increases faster than 
any other models.
This means that the accretion rate is highest among the models 
and leads to the shortest neutrino burst.  
Since the mass becomes large soon after bounce, 
the density and temperature inside the proto-neutron star 
become high and they cause the quick rise of average 
energies and luminosities as discussed above.  
This behavior is in accord with the analyses 
for their numerical results of the collapse of 40M$_{\odot}$ star 
by \citet{lie04}.  

\subsection{Profiles of density and temperature}\label{sec:profile}

We examine the profiles of three models during the evolution 
to discuss the origin of differences in the black hole formation 
and the neutrino signals.

\begin{figure}
\epsscale{0.7}
\plotone{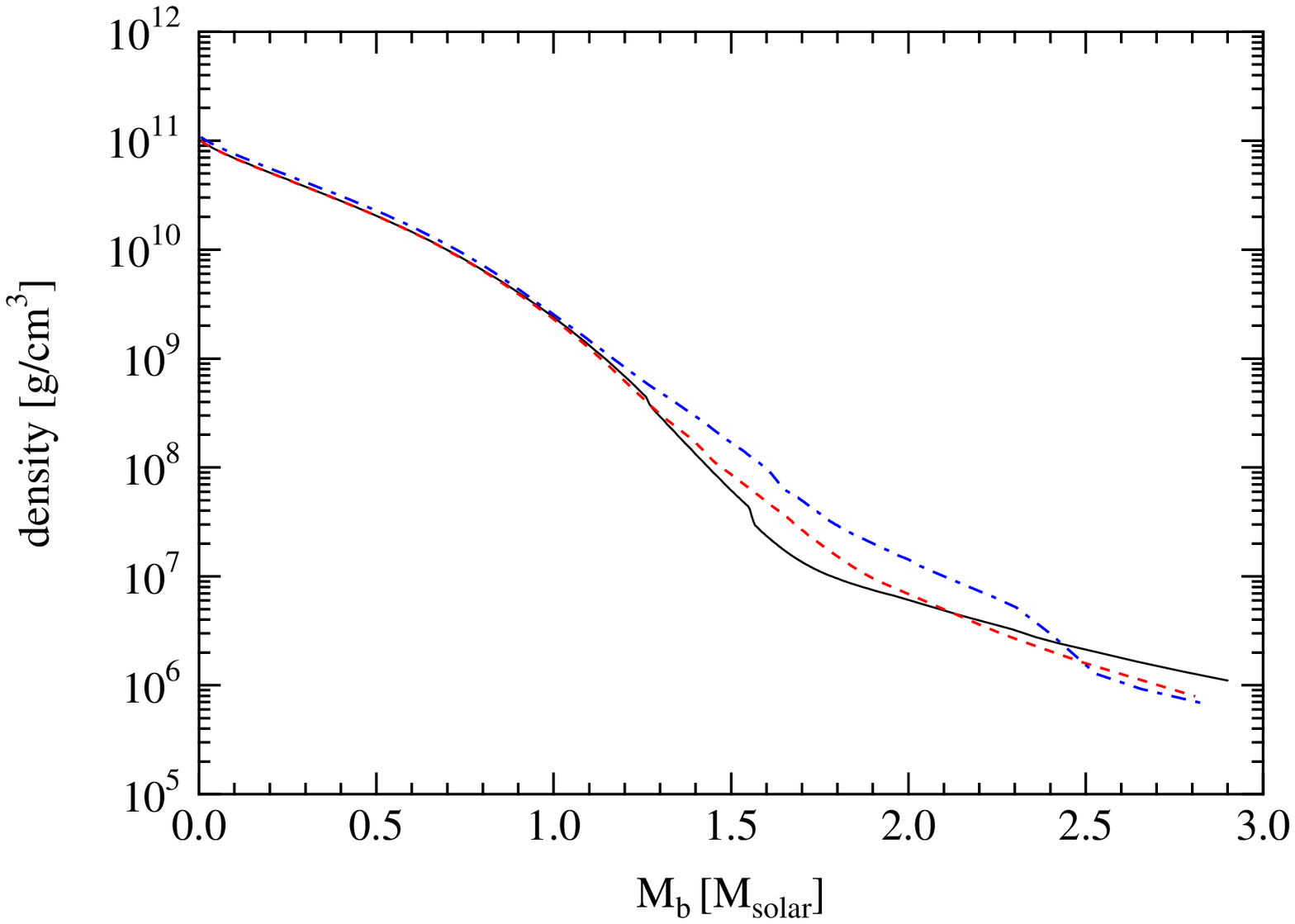}\\
\plotone{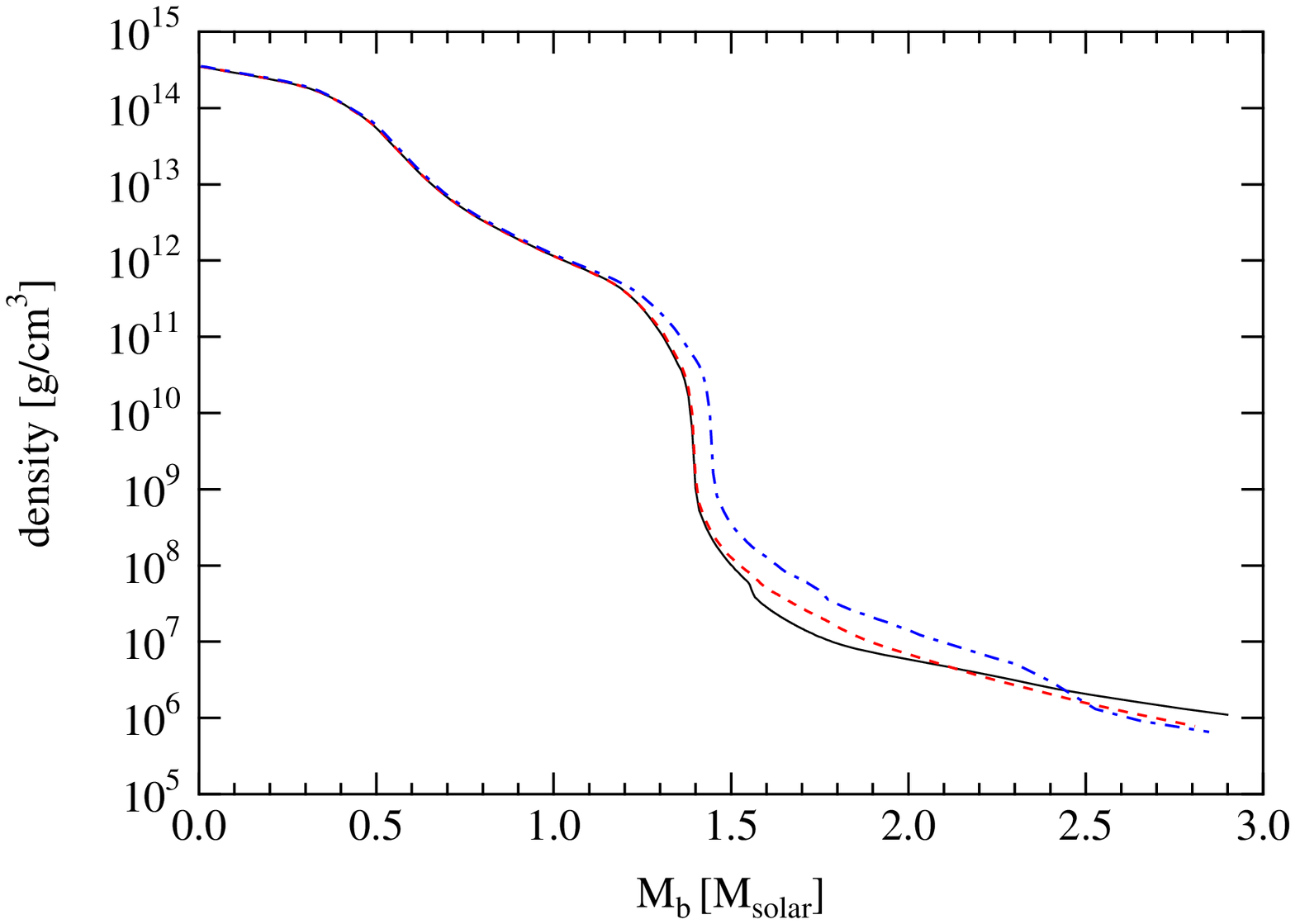}
\caption{Density distributions for W40L, T50L and H40L 
are shown as a function of baryon mass coordinate 
by solid, dashed and dash-dotted lines, respectively.
Upper and lower panels show the profiles 
at the time when the central density reaches 10$^{11}$ g/cm$^{3}$ 
and at 30 ms after bounce, respectively.}
\label{fig:profile-density}
\end{figure}

In order to clarify the origin of different histories of 
accretion rate, we make comparisons of density profiles 
for W40L, T50L and H40L 
at the time when the central density reaches 10$^{11}$ g/cm$^3$ 
and at 30 ms after bounce in Fig.~\ref{fig:profile-density}.  
In the upper panel of Fig.~\ref{fig:profile-density}, 
we see the same profiles at center (M$_b$ $\lesssim$ 1M$_{\odot}$) 
when we compare at the same central density during the collapse, 
though we have seen the differences among the initial models 
as a whole in Fig. \ref{fig:progenitor}.  
This similarity of central cores implies the similar dynamics 
around core bounce.  
In fact, the density profiles of central part 
(M$_b$ $\lesssim$ 1.2M$_{\odot}$) at 30 ms after bounce 
are quite similar as seen in the lower panel of 
Fig. \ref{fig:profile-density}.
We note that profiles of temperature and electron fraction 
are also similar in the central part.  

It is remarkable to see the difference of densities 
in outer part (M$_b$ $\gtrsim$ 1M$_{\odot}$) among the three models.  
Since the density at each point determines the time scale of free fall 
($\sim 1 / \sqrt{G \rho}$) 
onto the central object, 
the accretion rate ($\dot{M}=4\pi r^2 \rho v$) is 
proportional to $\sim \rho^{3/2}$, therefore, 
the density difference is the origin of different accretion rates.
In Fig.~\ref{fig:profile-density}, 
the higher density in the outer profile of H40L informs 
that the accretion rate is higher 
up to the point at M$_b$$\sim$2.4M$_{\odot}$, where there is 
a crossing with the two other models.  
The profiles of T50L and W40L are rather similar each other 
though there are slight differences and a crossing at M$_b$$\sim$2.1M$_{\odot}$.

\begin{figure}
\epsscale{0.7}
\plotone{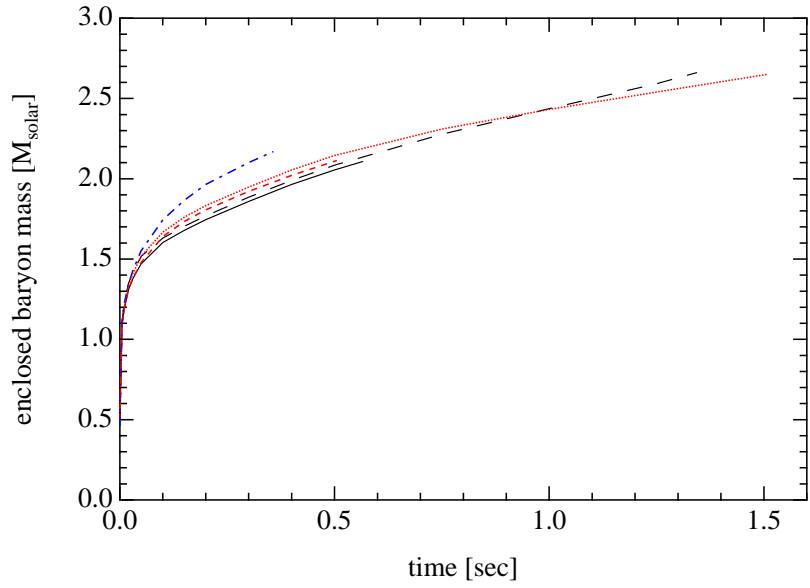}
\caption{Enclosed baryon mass in the central object 
(proto-neutron star) as a function of time after bounce 
for models W40S (long-dashed), W40L (solid), T50S (dotted), 
T50L (dashed) and H40L (dash-dotted).}
\label{fig:protoNSmass}
\end{figure}

The difference in the densities of outer layers remains 
during the evolution of proto-neutron star up to the black hole formation 
and leads to the different curves for proto-neutron star masses 
as we have seen in Fig.~\ref{fig:protoNSmass}.
At 30 ms after bounce, proto-neutron stars having $\sim$1.4M$_{\odot}$ 
(in baryon mass) are formed and the rest of outer layers 
are accreting onto them in the free fall time.  
As they evolve, 
the baryon mass of proto-neutron star of H40L becomes larger than 
those of W40L and T50L because of the higher accretion rate.  
The cases of W40L and T50L in Fig.~\ref{fig:protoNSmass}
are almost the same each other because of the similarity of density profiles.  
In the case of Shen-EOS, the cases of W40S and T50S 
in Fig.~\ref{fig:protoNSmass} behave similarly each other 
for the same reason.  
The increase in T50S at $\sim$1 s becomes slightly 
slower than that in W40S reflecting a crossing in the density 
profiles similar to the one discussed above.  

\begin{figure}
\epsscale{0.7}
\plotone{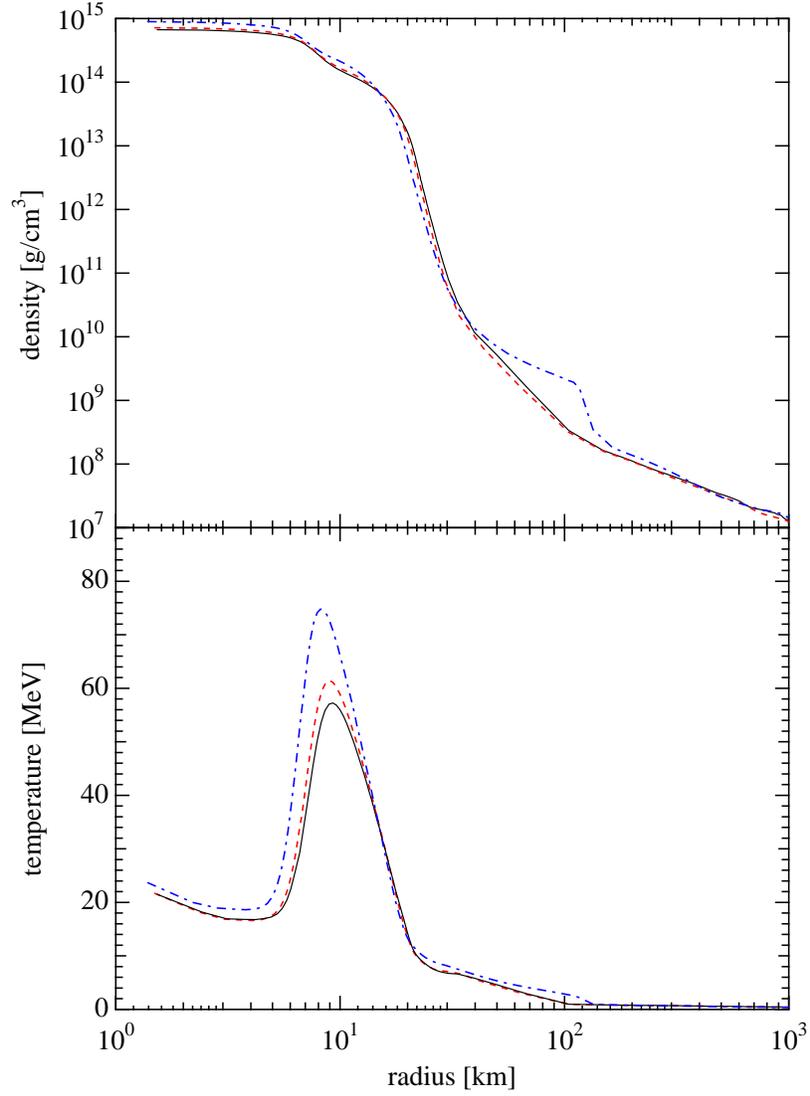}
\caption{Density (upper panel) and temperature (lower panel) 
distributions for W40L, T50L and H40L at 300 ms after bounce 
are shown as a function of radius 
by solid, dashed and dash-dotted lines, respectively.}
\label{fig:profile-tpb300}
\end{figure}

In Figure~\ref{fig:profile-tpb300}, we compare the profiles 
of density and temperature at 300 ms after bounce 
for W40L, T50L and H40L.  
The central density for H40L is higher than those for W40L and T50L 
due to the larger baryon mass in proto-neutron star for H40L 
as seen in Fig.~\ref{fig:protoNSmass}.  
The corresponding temperature for H40L is higher than others 
because of a more compact profile of proto-neutron star.  
We note that a kink at $\sim$100 km in the density profile of H40L 
corresponds to the shock position in accreting matter.  
The profiles of W40L and T50L are similar each other, 
though the profile in T50L is slightly more compact 
having higher density and temperature than those in W40L.  

The compact profile in H40L leads to the higher average 
energies and luminosities in neutrino emission than 
those in the two other models.  
The higher average energy of $\nu_{\mu/\tau}$ in H40L 
arises from the higher temperature in H40L than others.  
The higher luminosities in H40L is also caused by 
the higher temperature and the compactness of central 
object.  
Since the accretion luminosities of $\nu_e$ and $\bar{\nu}_e$ 
are determined by the liberated energy 
in the gravitational potential of central proto-neutron star, 
a more massive and compact object leads to larger luminosities.
In contrast, the neutrino energies and luminosities 
of W40L and T50L are similar each other because of the similar 
history of accretion and resulting evolution of proto-neutron 
stars.  
The same argument applies for the comparisons between W40S and T50S, 
in which the average energies are very close.  
The luminosities of T50S are somewhat smaller 
corresponding to the lower accretion rate at the late phase 
$\gtrsim$0.6 s after bounce.  

To summarize, the density profile in progenitors determines 
the accretion rate in black hole forming collapse and affects 
the evolution of proto-neutron stars toward the black hole formation.  
The difference in proto-neutron star profiles in turn leads to 
the different features in neutrino emission.  
The characteristics of neutrino emission reflect the minute differences 
in density profiles of progenitors while the influence 
from the equation of state is more significant.  


\subsection{Differences in neutrino signals}\label{sec:nu-signal}

Here we discuss the characteristics of neutrino bursts 
in the five calculated models 
aiming at discrimination of the effects from progenitors and EOSs.  
We summarize the numerical results regarding the black hole formation 
in Table~\ref{tb:model}.  

In order to probe the EOS, the time duration of neutrino burst is decisive.  
If the accretion rate of the progenitor can be determined, 
the measurement of the duration enables us to infer the critical 
mass of proto-neutron stars and to extract the stiffness of the EOS.  
In the current study, Shen-EOS is stiff and provides large critical 
masses of $\sim$2.7M$_{\odot}$ in baryon mass ($\sim$2.3M$_{\odot}$ 
in gravitational mass, see Table~\ref{tb:model}) for accreting 
proto-neutron stars.  
This large mass corresponds to the duration till the black hole 
formation over 1.3 s and is significantly longer than the cases 
of Lattimer-Swesty EOS as seen in Fig.~\ref{fig:protoNSmass}.
Since Lattimer-Swesty EOS is soft, the critical mass is 
$\sim$2.1M$_{\odot}$ in baryon mass ($\sim$2.0M$_{\odot}$ 
in gravitational mass).  
The duration is $\sim$0.5 s with weak 
dependence on the progenitors 
among T50L, H40L and W40L.  
If one can get the information of the time duration accurately, 
one can infer the critical mass through the calculated curves 
of mass increase.  
For example, if 
we assume that the time duration of neutrino burst is 0.3 s, 
the critical mass is in the range between 1.8M$_{\odot}$ 
and 2.1M$_{\odot}$ 
estimated 
from the range between W40L and H40L in Fig.~\ref{fig:protoNSmass}.  
Therefore, 
one can estimate the critical mass with an error of $\sim$0.3M$_{\odot}$ 
from the time duration 
within the uncertainty in the density profile of the progenitors.  
It is to be noted that the critical masses for proto-neutron stars, 
which contain abundant neutrinos at finite temperature, 
are different from the maximum masses for neutrino-less and cold 
neutron stars.  

To reveal the differences due to progenitors, it is necessary 
to examine the detailed information of average energies and 
luminosities.  
When Shen-EOS is adopted, one has to wait for $\sim$1 s 
to have enough differences between W40S and T50S.  
In case of Lattimer-Swesty EOS, it is rather difficult 
to distinguish the differences between the progenitors since 
the duration of burst is too short.  
The neutrino signals in W40L and T50L are too similar 
each other to discriminate.
H40L is different from W40L and T50L, on the other hand, 
and may be distinguishable if the data on the average energy 
of $\nu_{\mu/\tau}$ and/or accurate luminosities are available.  

\begin{figure}
\epsscale{0.7}
\plotone{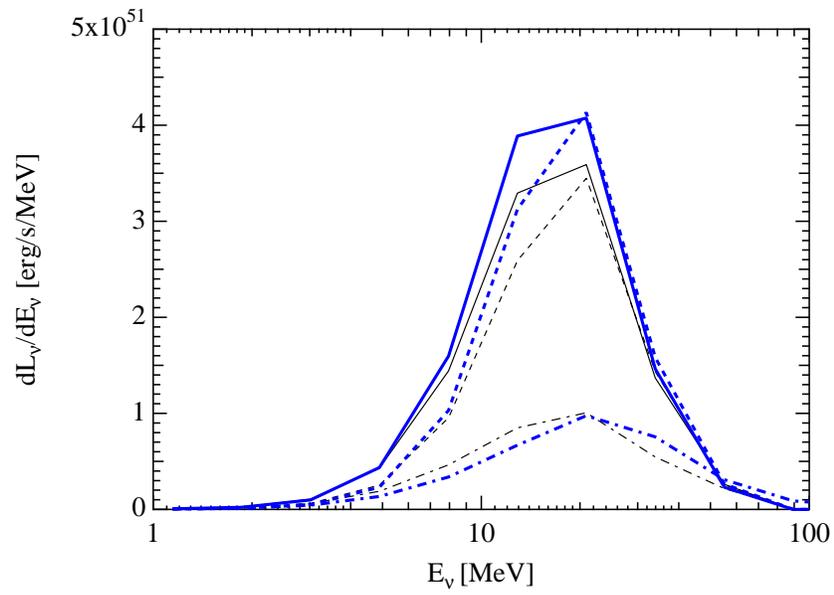}
\caption{Energy spectra at 300 ms after bounce are shown 
as a function of neutrino energy for 
$\nu_e$, $\bar{\nu}_e$ and $\nu_{\mu/\tau}$ 
by solid, dashed and dash-dotted lines, respectively, 
for model H40L (thick) and W40L (thin).}
\label{fig:spect-tpb300}
\end{figure}

In Figure~\ref{fig:spect-tpb300}, we compare the energy spectra of 
emitted neutrinos at 300 ms after bounce for H40L and W40L.
The shapes of energy spectra for $\nu_e$ and $\bar{\nu}_e$ are 
similar between H40L and W40L, however, the peak values are higher 
for H40L.  
The peak position for $\nu_{\mu/\tau}$ for H40L is shifted for a higher energy, 
though the peak value is close to that for W40L.  
These differences appear as higher luminosities of all species 
and higher average energy of $\nu_{\mu/\tau}$ 
(but similar average energies for $\nu_e$ and $\bar{\nu}_e$) for H40L 
as seen at 300 ms in Fig.\ref{fig:EnuLnu-H40L}.  
We compared the energy spectra at 300 ms after bounce 
also for T50L and W40L (not shown here).  
The energy spectra are similar each other except for slightly 
higher peak values for $\nu_e$ and $\bar{\nu}_e$ for T50L.  

We remark here that the discussions on neutrino signals so far 
do not take into account the neutrino oscillations.  
One should take into account the changes of spectra 
during the propagation of neutrinos in outer layers.  
The observational aspects of the current numerical results 
considering the effects of neutrino oscillations 
are now under investigation and will be published elsewhere.  

\section{Discussions}\label{sec:discussions}

In this section, we discuss 
the perspective of the current work 
regarding the progenitor and the equation of state.  
We also remark on 
the observational chances of neutrino burst 
and implications to astrophysical phenomena 
with the black hole.  

The present study on 
the black hole formation and the associated neutrino burst 
relies on the presupernova configurations of massive stars.  
Further studies with updated models of progenitors 
are definitely warranted.  
Necessary condition to have this type of phenomena is 
the appropriate density distribution to cause the intense accretion.  
In the three models (T50, H40 and W40) adopted in the current study, 
the size of iron core is large ($\sim$2M$_{\odot}$) 
and the dense region 
extends up to M$_b$ $\sim$3M$_{\odot}$. 
This is different from the case of a 15M$_{\odot}$ star 
by \citet{woo95}, in which the iron core is smaller and the density 
slope is steep already at M$_b$ $\sim$1.5M$_{\odot}$, for example.
It is necessary to study further the density profiles 
of various progenitors and its relation to the accretion 
in order to examine whether the size of iron core is essential.  
It may be enough to have the flat profile of density 
so that the density remains high ($\sim$10$^{6}$ g/cm$^{3}$) 
at M$_b$=2--3M$_{\odot}$, which corresponds to the critical 
mass of proto-neutron stars.  
If the density profile is steep and the density is too low 
at the corresponding region in 40--50M$_{\odot}$ stars, 
the black hole formation may take much longer ($\sim$10--100 s) 
than the current case.  
Such events with longer neutrino bursts may look similar to 
the ordinary supernovae of $\sim$15M$_{\odot}$ stars in terms of 
neutrino signals.  

Regarding the uncertainty of progenitors, one should 
pay attention to the mass loss as well as the rotation.  
The progenitor models in \citet{woo95} and \citet{has95} are obtained 
by the presupernova evolution without the mass loss.  
The inclusion of the mass loss leads to the smaller 
mass at the precollapse stage even from the large 
initial mass (ex. 50M$_{\odot}$) at the main sequence.  
The iron core mass and the density distribution may 
end up to be similar to those in the progenitors 
of about 20M$_{\odot}$.  
In this regard, metal poor massive stars are more 
likely to be the progenitors for the black hole formation 
as studied in the current work.  
In fact, T50 adopted in the current study is 
the one with zero metallicity that forbids the mass loss.
Its large central core results in the necessary intense accretion, 
which in turn leads to the quick black hole formation 
with short neutrino bursts.  
Whether the change of evolution toward the solar metallicity 
may cause the suppression of the current scenario 
remains to be seen along with the systematic 
studies on the metallicity dependence of progenitors 
and the careful studies on the mass loss.  

We study the black hole formation under the spherical symmetry.
This geometrical assumption limits ourselves to the non-rotating 
massive stars among varieties in rotation rates.
One should note, however, that the spherical symmetry is a good approximation 
if the rotation is not significantly fast.  
It is challenging to study the collapse of rapidly rotating 
massive stars, which are possible progenitors for hypernovae 
and gamma ray bursts.  
One has to perform the fully general relativistic calculations 
in multi-dimensions 
to follow the evolution toward the black hole formation.  
Such challenges are made in the numerical simulations of 
general relativistic hydrodynamics but with simple treatment 
of neutrino transfer, microphysics (ex. EOS) and 
progenitors \citep{shi05,sek05,sek07}.  
Multi-dimensional simulations with proper treatment of 
neutrino transfer are awaited for to predict the features 
of neutrino emission.  

The rotating collapse to the black hole and the following 
evolution is also interesting from the aspect of 
nucleosynthesis.  
In the context of collapsar scenario, the ejection of 
neutron-rich material in the jet from the surroundings 
of black hole and the neutrino-driven winds from the 
accretion disk around the black hole are proposed to 
be the possible site for the heavy elements production 
including r-process elements \citep{sur06,fuj07}.  
Although we performed the numerical simulations 
in the limit of no rotation, one may use the current results 
to guess the time duration till the black hole formation 
and the accretion rate of material, 
which are influential in the scenario of nucleosynthesis.  
The time duration obtained in the current work is a lower limit 
since the maximum mass of rotating proto-neutron stars 
is larger than the non-rotating ones \citep{sum99} 
and the rotation supports 
accreting material away due to the centrifugal force.  
The accretion rate in the current study is an upper limit, 
accordingly.  
Large accretion rates in the disk formed around black hole, 
which are assumed in the studies of nucleosynthesis 
of r-process \citep{sur06}, 
may be limited from the density profile of progenitors 
as we have pointed out.  

The observational possibility to detect the neutrino bursts 
toward the black hole formation from massive stars deserves 
to be mentioned.  
Occurrence rate of the phenomena (non-rotating black hole formation 
and its short neutrino burst) as in the current scenario 
is proportional to the fraction of massive stars having 
large masses for intense accretion without significant rotation.  
This depends on the initial mass function and rotation rates 
with large uncertainties, but such stars comprise a certain fraction 
($\sim$30$\%$) of massive stars, which can be estimated from the 
Salpeter's initial mass function with an assumption on 
the mass threshold ($\sim 25$M$_{\odot}$) for the black hole formation.  
\citet{koc08} estimated from the observations done so far that 
the rate of black hole formation could be comparable to the rate of 
normal core-collapse supernovae.  
Observational analyses of explosion energies and Ni productions 
for various supernovae 
indicate that there 
have been observed at least several 
faint supernovae 
(SN~1997D and SN~1999br, for example) 
from massive stars beyond 20M$_{\odot}$ without significant 
rotations, which may be akin to the event considered in this 
paper \citep{nom07}.  

The planned survey of disappearance of massive stars 
within a distance of 10 Mpc \citep{koc08} 
will find both dim supernovae and collapses with no optical display 
that lead to the black hole formation.  
If they occur in the Galaxy or nearby galaxies, the information on progenitor 
will be obtained by the optical records before it disappeared.  
In addition, thousands of neutrinos will be detected 
at the terrestrial facilities.  
Neutrino data combined with optical ones will  
provide us with the information on the dynamics of black hole
formation and constrain the properties of dense matter and 
the progenitor.  
In the search of supernova neutrinos by the Super-Kamiokande detector, 
neither neutrinos associated with black hole formation 
nor ordinary supernova neutrinos are detected 
during the period of their survey unfortunately \citep{ike07}.  

The numerical result on the massive star with zero metallicity (T50) 
suggests that the black hole formation in the current scenario 
may occur for Pop III stars in the early Universe 
(see also \citet{nak06,nak07}).  
Since the massive stars become more popular in the metal poor epoch, 
the neutrinos from black hole formation in the early stage of galaxies 
may contribute to the relic signal as back ground neutrinos.
Although the signal is short, higher energies and luminosities 
may help to enhance this contribution and they may not be negligible.  
This argument again depends on the fraction of massive stars 
resulting the current scenario through the evolution of galaxies.  

Before we close the discussion, we comment on the choice 
of equation of state.  
Although density and temperature become extremely high 
enough to have new degrees of freedom such as hyperons 
and quarks in the evolution toward the black hole formation,
we adopt the two sets of EOS within the nucleonic 
degree of freedom in the current study to assess 
the progenitor dependence.  
Studies on the modifications of the evolution of black hole 
formation and the neutrino signal due to hyperon mixtures
and the appearance of quark phase are now under way and 
will be reported in separate papers 
\citep{nak08}.  

\section{Summary}\label{sec:summary}

We study the black hole formation and the associated neutrino emission 
from the gravitational collapse of 
massive stars to clarify the dependence on the progenitor models 
following the basic study of the phenomena in the first paper \citep{sum07} 
of the series.
We perform the sets of numerical simulations of general relativistic 
$\nu$-radiation hydrodynamics under the spherical symmetry to follow 
the collapse of massive stars of 40M$_{\odot}$ and 50M$_{\odot}$ 
adopting different models of the presupernova evolution. 
We clarify that the characteristics of evolution of proto-neutron stars 
toward the black hole formation and 
the associated short burst of neutrinos are found similar 
in the numerical simulations for the three progenitor models.  

We find that the different progenitor models may result in different rates 
of matter accretion onto the proto-neutron star born in the collapse 
of massive stars.  
The different accretion rates can lead to different histories of increasing 
mass of proto-neutron stars toward the critical mass and to different durations till the black hole formation.  
In order to extract the properties of equation of state 
from the neutrino burst toward the black hole formation 
from massive stars, 
as proposed in the previous studies \citep{sum06,sum07}, 
it is necessary to constrain the uncertainty from progenitor models.  
In the comparison of the three progenitors we studied, 
the change of time duration is found not so large as  
the difference originating from the two sets of equation of state.  
The change of time profiles of average energies and luminosities of 
neutrinos due to progenitors is found to be minor as well. 
Note, however, that the present study is a first trial exploration and 
more systematic investigations are obviously necessary to see 
if our conclusion is generic or not.  

We demonstrate that there is a relation between 
the density profile of progenitor and the accretion history 
to cause the difference in black hole formation.  
A higher density at M$_{b}$=2.0--3.0M$_{\odot}$ in a progenitor 
results in a shorter free fall time and a higher accretion rate.  
A flat density profile of core in massive stars is a key factor 
to cause the rapid formation of black hole formation in $\sim$1 s 
together with the characteristic neutrino burst.  
When the density gradients are different as in the comparison 
between W40L and H40L, the density profiles may be distinguished 
from the detailed information of neutrino detections.  
This relation of the density profile of progenitors 
to the accretion rates is helpful to discuss the evolution 
after the collapse of massive stars in other stellar models 
and may be useful to infer the accretion dynamics of rotating 
collapse to the black hole formation and associated explosive 
phenomena such as collapsars.  
Further studies on rotating collapse of massive stars 
with intense accretion are called for to clarify whether 
the outcome may lead to possible success of modeling 
gamma ray burst and/or nucleosynthesis of heavy elements.  

\acknowledgments
The authors are grateful to H. Shen, K. Oyamatsu, 
A. Ohnishi, C. Ishizuka and H. Toki 
for the collaborations on the tables of equation of state 
for supernova simulations.  
We thank K. Nakazato, S. Fujimoto, K. Kotake, M. Liebendorfer 
and T. Kajino for fruitful discussions and keen comments.
We thank N. Tominaga, H. Umeda, K. Nomoto and M. Hashimoto 
for providing their progenitor models and further correspondence.  
K. S. expresses his gratitude to W. Hillebrandt and T.-H. Janka 
for the hospitality at Max Planck Institute f\"ur Astrophysik 
where a part of this work was done.  
The numerical simulations were performed on VPP5000 
at Center for Computational Astrophysics, CfCA, of 
National Astronomical Observatory of Japan (wks06a, iks13a, uks06a), 
on Altix3700Bx2 at the Tokai branch of Japan Atomic Energy Agency (JAEA), 
on Altix3700Bx2 at YITP in Kyoto University 
and SX at RCNP, Osaka University.  
K. S. is grateful to T. Maruyama and S. Chiba 
for the arrangement of computing resources at JAEA.  
This work is partially supported by the Grants-in-Aid for the 
Scientific Research (18540291, 18540295, 19104006, 19540252) 
of the MEXT of Japan, Academic Frontier Project of the MEXT, KEK LSSP (07-05)
and the 21st-Century COE Program "Holistic Research and Education 
Center for Physics of Self-organization Systems".

\clearpage
\begin{table}
\begin{center}
\caption{Summary of calculated models\label{tb:model}}
\begin{tabular}{cccccccc} \tableline\tableline
Model & Progenitor                  & M$_{prog}$ & M$_{Fe}$ & EOS  & M$^{max}_{b}$ [M$_{\odot}$] & M$^{max}_{g}$ [M$_{\odot}$] & t$_{BH}$ [s]  \\ 
\tableline
W40S  & WW95\tablenotemark{a}       & 40         & 1.98     & Shen & 2.66                        & 2.38                        & 1.35            \\
W40L  & WW95                        & 40         & 1.98     & LS   & 2.10                        & 1.99                        & 0.57            \\
T50S  & TUN07\tablenotemark{b}      & 50         & 1.88     & Shen & 2.65                        & 2.33                        & 1.51            \\
T50L  & TUN07                       & 50         & 1.88     & LS   & 2.11                        & 2.01                        & 0.51            \\
H40L  & H95\tablenotemark{c}        & 40         & 1.88     & LS   & 2.17                        & 2.08                        & 0.36            \\ 
\tableline
\end{tabular}
\tablenotetext{a}{WW95:~\citet{woo95}}
\tablenotetext{b}{TUN07:~\citet{ume05,tom07}}
\tablenotetext{c}{H95:~\citet{has95}}
\tablecomments{M$_{prog}$ and M$_{Fe}$ is the mass of progenitor and the iron core mass, respectively.  
M$^{max}_{b}$ and M$^{max}_{g}$ are the masses of proto-neutron star 
just prior to the re-collapse in baryon and gravitational masses, respectively.  
t$_{BH}$ is the time at the black hole formation measured from the core bounce.  
See the main text for details.  }
\end{center}
\end{table}

\clearpage

\end{document}